\documentclass[twocolumn,superscriptaddress,showpacs,floatfix,amsmath,amssymb,aps,pre,longbibliography]{revtex4-1}

\usepackage{hyperref}
\usepackage[usenames,dvipsnames,svgnames,table]{xcolor}
\definecolor{PlotBlue}{HTML}{1F77B4}
\definecolor{PlotOrange}{HTML}{FF7F0E}
\definecolor{PlotGreen}{HTML}{2CA02C}
\definecolor{PlotRed}{HTML}{FF0000}
\usepackage{graphicx}

\begin{document}
	
\title{Equivalence of nonequilibrium ensembles in turbulence models}
\thanks{Version accepted for publication (postprint) on Phys. Rev. E 98, 012202 -- Published 3 July 2018}

\author{Luca Biferale}
\affiliation{Dipartimento di Fisica and INFN, Universit\`a di Roma ``Tor Vergata'', Via Ricerca Scientifica 1, 00133 Roma, Italy}

\author{Massimo Cencini}\affiliation{Istituto dei Sistemi
	Complessi, CNR, via dei Taurini 19, 00185 Rome, Italy and INFN ``Tor Vergata''}

\author{Massimo De Pietro}
\affiliation{Dipartimento di Fisica and INFN, Universit\`a di Roma ``Tor Vergata'', Via Ricerca Scientifica 1, 00133 Roma, Italy}

\author{Giovanni Gallavotti}
\affiliation{INFN-Roma1 and Universit\`a "La Sapienza” Roma, Italy}

\author{Valerio Lucarini}
\affiliation{ Department of Mathematics and Statistics, University of Reading, Reading, RG66AX, United Kingdom}
\affiliation{ Centre for the Mathematics of Planet Earth, University of Reading, Reading, RG66AX, United Kingdom}
\affiliation{  CEN, University of Hamburg, Hamburg, 20144, Germany}

\date{\today}

\begin{abstract}
	Understanding under what conditions it is possible to construct 
	equivalent ensembles is key to advancing our ability to connect 
	microscopic and macroscopic properties of non-equilibrium statistical 
	mechanics. In the case of fluid dynamical systems, one issue is to 
	test whether different models for viscosity lead to the same 
	macroscopic properties of the fluid systems in different regimes. Such 
	models include, besides the standard choice of constant viscosity,  
	cases where the time symmetry of the evolution equations is exactly 
	preserved, as it must be in the corresponding microscopic systems, when 
	available. Here a time-reversible dynamics  is obtained by imposing the 
	conservation of global observables. We test the equivalence of 
	reversible and irreversible  ensembles for the case of a multiscale 
	shell model of turbulence. We verify that the equivalence is obeyed for 
	the mean values of macroscopic observables, up to an error 
	that vanishes as the system becomes more and more chaotic. 
\end{abstract}
\maketitle

\section{Introduction}
The macroscopic description of the dynamics of physical systems 
typically include forces that phenomenologically model the effect of 
molecular disordered motions, and are controlled by appropriate 
transport coefficients (such as viscosity, diffusivity,s etc.). A 
prominent example is given by the viscous term of the Navier-Stokes (NS)
equations. Such forces break the time reversibility, which is instead 
inherent in the microscopic dynamics. They are also responsible for the 
dissipation of energy, which allows for establishing  a 
(non-equilibrium) statistically steady state when the system is 
externally driven.

In the context of molecular dynamics, a similar role is played by
thermostats. A body of numerical simulations have shown that the non-equilibrium properties of systems
composed of a large number of molecules (particles) are basically
independent of the precise nature (reversible or not) of the model
used for the thermostats \cite{morriss2013statistical}. This suggests
that something similar may apply to the macroscopic description of
physical systems, as pioneered in simulations in
\cite{she1993constrained} and conjectured, on more theoretical grounds,
about two decades ago in
\cite{gallavotti1996equivalence,Gallavotti1997Dynamical}. Specifically, 
the hypothesis is that the statistical properties of the
non equilibrium steady state of a macroscopic system, whose dynamics
obeys a simple phenomenological law of the kind described above,
should be equivalently described by different macroscopic equations,
including some that preserve time-reversal symmetry. In
particular, with the example of fluid dynamics in mind, this can be
realized by allowing the viscosity to depend on the fluid velocity in
an appropriate way, thus converting the (inherently irreversible)
dynamical ensemble of the Navier-Stokes equations with a fixed
viscosity into a (formally reversible) dynamical ensemble with
fluctuating viscosity. In systems at equilibrium, a conceptually
similar step is done when switching from the microcanonical to the
canonical ensemble.

The equivalence discussed above has already been scrutinized in a few
simple systems such as a highly truncated version of the two-dimensional (2D) Navier-Stokes equations with periodic boundary conditions
\cite{rondoni1999fluctuations,gallavotti2004lyapunov} and more
recently in the Lorenz model \cite{gallavotti2014equivalence}.
Such tests dealt with systems not exhibiting the timescale separation
typical of many macroscopic systems. In this paper we explore whether (and under what conditions)
it is possible to establish an equivalence
between different nonequilibrium ensembles in systems with multiple
spatial and temporal scales. In particular, we investigate the shell
model for turbulence introduced in
\cite{Lvov_1998_improved_shellmodels} (see also
\cite{biferale2003shell,bohr2005dynamical} for general surveys on
shell models).

The study of multiscale systems is at the core of many disciplines
dealing with complex systems and the construction of accurate methods
for model reduction is of great relevance for the theory and for the
construction of efficient and robust numerical models. For instance a
substantial part of the effort in weather and climate modeling is
devoted to improving the representation of small-scale processes.
This requires a difficult interplay between large eddy simulations
(LESs) \cite{galperin1993large,sagaut2006large} and dedicated
observational campaigns. Large-eddy simulations themselves
need to be tailored via parametrization, which amounts to defining
suitable subgrid models, the so-called eddy viscosities, to be
compatible with direct numerical simulations (DNSs) of turbulent
flows. The term to be modeled in LESs is
intrinsically time reversible, as derived from filtering the
nonlinear term of the Navier-Stokes equations.  Indeed,
reversible eddy viscosity models have been studied in
\cite{carati2001modelling,fang2012time,Jimenez2015}.  However,
reversible subgrid models are prone to dynamical instabilities that
are often cured by adding suitable dissipative regularizations. As
it will be clear later, one of the remarkable properties of the
reversible dynamical systems studied in this paper is that it allows us to devise a
viscous modeling that binds the system to evolve in a finite region of the 
phase space.

A further motivation for studying reversible models for the dissipation is the possibility of employing
the universality properties known for the fluctuations of the
dissipation in reversible systems to infer, via the proposed
equivalence and the chaotic hypothesis in
\cite{gallavotti1995dynamical}, the validity of the same properties
in the standard irreversible model (i.e. shell model or Navier-Stokes
equations). For instance the fluctuation relation could be tested even
in the irreversible evolutions \cite{gallavotti2017reversible}. The observed
pairing (to a nonconstant line \cite{Ga997b,gallavotti2014equivalence}) empirically observed in a few
simulations on the 2D Navier-Stokes equations, reversible and
irreversible \cite{gallavotti2004lyapunov}, could lead to a precise
determination of the ``slope'' of the fluctuation relations, if confirmed
by dedicated simulations. Further, the equivalence conjecture could in principle be
used for prediction of local fluctuations of dissipation in standard NS
evolutions.

The paper is structured as follows. In Sec. \ref{sec:equivalence} we 
provide a concise, but self-contained, summary of the general framework 
of nonequilibrium dynamical ensemble equivalence,  where we formalize a 
general theoretical approach to the problem in the form of conjectures 
that can be subjected to tests.

In Sec. \ref{sec:models} we present the multiscale model analyzed in 
detail in this study.  We supplement the traditional irreversible model 
with the reversible models obtained by replacing viscous terms with 
forces imposing anholonomous constraints on suitably chosen observables 
selected so that the resulting equations are time reversible.

Comparisons between properties of the irreversible and reversible 
models are discussed in Sec. \ref{sec:numerical}, where we analyze a 
range of mathematical and physical properties of the models and assess 
whether the equivalence discussed above holds.

In Sec. \ref{sec:conclusions} we summarize and discuss the main 
findings of our paper and present perspectives for future works in this 
direction.

\section{The General Framework: Equivalence of 
	Ensembles\label{sec:equivalence}} 
\subsection{Equivalence of Equilibrium 
	Ensembles\label{sec:equivalence_intro}} 
One of the cornerstones of equilibrium statistical mechanics is the
possibility of establishing an equivalence between different
statistical ensembles \cite{touchette2011,touchette2015}. This means
that in the thermodynamic limit, as the number of particles goes to
infinity, the expectation values of physical observables of the
system do not depend on the specific choice of the thermostat defining
the interaction between the system and the reservoir it is in contact
with, when suitable consistency is imposed.

Clearly not all physical observables will have the same value
in the different ensembles. For instance, in a system statistically
described by the canonical ensemble the temperature fluctuations
vanish while energy fluctuates and the opposite occurs in a system
described by the microcanonical ensemble.

The equivalence of equilibrium ensembles allows us to understand the 
emergence of macroscopic thermodynamical properties that do 
not depend on the details of the microscopic dynamics describing the 
coupling between a system and the surrounding environment.

\subsection{Equivalence of Non-Equilibrium Ensembles\label{sec:gen-equivalence_theo}
}
\subsubsection{General Discussion}

Let us consider the simplest case of an out-of-equilibrium system 
modeled by a differential equation with  $N$ variables that can be 
thought of as a time-reversible equation perturbed with an external force, 
which injects energy into the system, plus a dissipative force, which 
absorbs energy, allowing the system to reach a steady state.

The parameter controlling dissipation (e.g. viscosity $\nu$ in a 
fluid) can be replaced by a multiplier  defined in such a way that the 
new equation admits a suitably selected observable as an exact constant 
of motion (e.g. the fluid enstrophy). We will call it the balancing observable. 

Furthermore the multiplier can often be chosen so that the new
equations exhibit a time reversal symmetry (see below for typical
examples). The multiplier will fluctuate in time and for
macroscopic observables an equivalence is expected between the
irreversible and reversible formulations. By macroscopic we
mean observables that depend on a few (much less than $N$) large-scale degrees of
freedom (hence insensitive to the details of the system when $N$ is
large). One expects the equivalence to hold when the motion is
sufficiently chaotic and the fluctuating multiplier has an average equal
to the value of the phenomenological dissipation parameter.

For instance, in the case of a fluid described by the incompressible 
Navier-Stokes equations  in a homogeneous geometry (e.g. periodic 
boundary conditions)  or by a shell model truncated at $N$  modes we 
have a multiscale nonequilibrium system characterized by a single 
dynamical parameter $R$, the Reynolds number, and an ultraviolet cutoff 
$N$. At fixed forcing, equivalence of the averages of a prefixed number 
of observable  is expected in the limit of very small dissipation, e.g.,   
$\nu\to0$ or, equivalently, Reynolds number $R\to\infty$. The discrepancy between 
averages of the prefixed observables is expected to become smaller than 
some $\delta>0$ for an $R$ above a threshold value $R_{\delta}$. In 
applications to fluids it is also expected that $N$  should be taken 
large enough and correspondingly the equivalence threshold 
will have to become $R>R_{\delta,N}$, i.e. depending on $N$ 
too. The order of the limit $R \to \infty$ and $N \to \infty$ is a 
delicate issue that will be discussed in the following for the specific 
case of the shell model. 

Then the analogy with the usual theory of ensemble equivalence for 
equilibrium statistical mechanics would be complete with $\nu$, or the 
Reynolds number $R \propto \nu^{-1}$, playing the role of the inverse 
temperature and with  $N$  (necessary, perhaps, to give  mathematical 
wellposedness to the equations in three dimensions or certainly in  
numerical implementations in any dimension) playing the role of the 
volume. 

Of course, an important question is how we choose the
balancing observable in order to successfully define an
equivalent ensemble For instance, in the NS case, the balancing can
be constructed using the total enstrophy, or the total energy, or
other macroscopic observables. The choice might be critical because
the Fourier components of the velocity field have non-local
interactions \cite{kraichnan1976eddy}, so the equivalence could
be affected by the same difficulties that occur in equilibrium
statistical mechanics in systems with long-range interactions
\cite{leyvraz2002ensemble}. There is no general
prescription and, in the end, the choice might be based on empirical
grounds or motivated by or targeted to specific applications.

\subsubsection{Mathematical Formulation\label{sec:equivalence_theo}} 
\label{sec:equiv_conj_statement} 
Mathematically speaking, we consider a dynamical system with $N$ 
degrees of freedom written as
\begin{equation}
\dot x_j= f_j(x)+ F_j- \nu (\mathbb{L} x)_j\,, \quad j=1,\ldots,N\,,
\label{e1.1a}
\end{equation}
where $F_j$ is a constant forcing, $\nu>0$ is a dissipation
coefficient, and $\mathbb{L}$ is a positive-definite dissipation
matrix. In many interesting cases one has $(\mathbb{L}x)_j = g_j x_j$
with $g_j>0$ (the matrix is diagonal and all the elements on the
diagonal are positive; no summation is implied here).

A system is said to have a time-reversal symmetry $I$ if the map $I$ 
acts on the variable $x$ so that if $t\to S_t x$ is a solution then $I 
S_t x = S_{-t} I x$, i.e. if $t\to x(t)$ is a solution also $I x(-t)$ is 
solution [with the datum $I x(0)$)]. Therefore, Eq. (\ref{e1.1a}) has 
the map $Ix=-x$ as a time-reversal symmetry if $f_j(x)$ is even in $x$ and 
$\nu=0$.

Let $O(x)$ be an observable such that $\sum_{j=1}^N \partial_j O(x)
(\mathbb{L}x)_j=M(x)$ is positive for $x\ne0$. For instance, in Eq. (\ref{e1.1a})
if  $L = \mathbb{I} $ and $O(x) = x^2 / 2$ then 
$M(x)= x^2$. Then  the equation
\begin{equation}
\dot x_j = f_j(x)+ F_j- \alpha(x) (\mathbb{L} x)_j \, ,
\label{e1.2a}
\end{equation}
with
\begin{equation}
\alpha(x) \equiv \frac{\sum_{j=1}^N (f_j+F_j) x_j}{M(x)}\, ,   \label{e1.3a}
\end{equation}
admits $O(x)$ as an exact constant of motion, i.e., $\dot{O}=0$.
Furthermore, if $O(x)=O(-x)$, the equation is time reversible. In Eq.
(\ref{e1.1a}) the viscosity $\nu$ is set constant and the observable
$O$ fluctuates; correspondingly, in Eq. (\ref{e1.2a}), the observable
$O$ is constant and the ``viscosity'' $\alpha$ fluctuates.

Hereafter the notation $X |_y$ will denote that $X$ is evaluated in the 
model where the quantity $y$ is kept constant. We say that the 
stationary distributions of Eqs. (\ref{e1.1a}) and (\ref{e1.2a})  
define ensembles of statistical distributions that can be 
parametrized by the value of $\nu$ for Eq. (\ref{e1.1a}) or by the 
(constant) value $\tilde{O}$ of the  observable $O$ for Eq. 
(\ref{e1.2a}).

In this work, equivalence means that the reflexivity property 
holds, i.e.
\begin{equation}
\langle O \rangle |_{\nu} = \widetilde{O} \,  \leftrightarrow \, 
\langle \alpha \rangle |_{\widetilde O}= \nu \, ,
\label{e1.4a}
\end{equation}
and for a given set of macroscopic observables $\Phi$ the stationary 
averages in the reversible and irreversible  evolutions are related by
\begin{equation}
\langle \Phi \rangle |_{\nu} = \langle \Phi \rangle |_{\widetilde O}\, 
(1+o)\, ,
\label{eq:conjecture1}
\end{equation}
with $o$ a $\Phi$-dependent quantity, infinitesimal  as 
$\nu^{-1}\to\infty$, for fixed $N$.

Equation (\ref{eq:conjecture1}) clarifies the thermodynamical aspect of 
the equivalence: In a strongly chaotic regime, measuring a macroscopic 
observable of the system, we are unable to say whether we are observing 
the reversible or the irreversible variant.

The property of reflexivity is an essential element of the proposed 
equivalence: Setting the value of the viscosity coefficient $\nu$ in 
the irreversible system is conceptually equivalent to setting the value 
of the physical quantity $O$ in the corresponding reversible system.

The above-mentioned formulation of the  equivalence conjecture
has been extended in other studies
\cite{gallavotti1995dynamical,gallavotti1995reversible,Ga997b,GRS004},
including the definition of a fluctuation relation for the reversible
ensemble, as well as conjectures about the equivalence of the Lyapunov
spectra in the two ensembles
\cite{gallavotti2014equivalence}. However, these concepts are beyond
the scope of this paper and will not be discussed in the following.

Finally, by repeating the procedure described above with a different 
observable $O$, it is possible to generate different time-reversible 
models  so that a plurality of (potentially equivalent) non equilibrium 
ensembles can in principle be constructed. 

\section{Models\label{sec:models}}

\subsection{The (irreversible) Shell Model\label{subsec:models_shell}} 
Shell models are finite-dimensional  chaotic dynamical systems 
providing a test bed for fundamental studies of fully developed 
turbulence \cite{frish_turbulence,bohr2005dynamical,biferale2003shell}. 
They can be thought of as  drastic simplifications of the Navier-Stokes 
equations and share with them many non trivial properties observed in 
experiments and simulations, such as the energy cascade from large to 
small scales, dissipative anomaly, and intermittency with anomalous 
scaling for the velocity statistics.  

Our analysis is based on the shell model introduced in Ref. 
\cite{Lvov_1998_improved_shellmodels}. It describes the evolution of a 
set of complex variables $u_n$, representing the velocity in a shell of 
wave numbers $ |\mathbf{k}| \in [k_n,k_{n+1}]$, with $n=0,\ldots,N-1$. 
The Fourier shells $k_n$ are geometrically spaced, $k_n=k_0 2^n$  with 
$k_0=1$, so that a large $O(2^{N})$ range of scales can be explored 
using few degrees of freedom. The equations of motion take the form 
\cite{Lvov_1998_improved_shellmodels}
\begin{equation}
\label{eq:sabra_standard_equations}
\dot{u}_n = \mathcal{N}[\{u_n\}] - \nu k_n^2 u_n + F_n \, , \quad n=0,
\ldots,N-1 \, ,
\end{equation}
where
\begin{equation}
\label{eq:sabra_standard_NLT}
\mathcal{N}[\{u_n\}] = i k_n \left(2 a u_{n+2} u_{n+1}^{*} \!+\! b 
u_{n+1} u_{n-1}^{*} \!+\! \frac{c}{2} u_{n-1} u_{n-2}\right) \, 
\end{equation}
accounts for the non linear coupling between neighboring wave numbers, 
$ -\nu k_n^2 $ is the dissipative term, and $F_n$ is an external 
force typically acting at large scales (here $F_n=F 
\delta_{n,0}$, with $F$ constant). The boundary conditions 
$u_{-1}=u_{-2}= u_{N}=u_{N+1}=0$ are imposed.

Rigorous results \cite{CLT007} have been derived for Eq. 
(\ref{eq:sabra_standard_equations}), proving that it admits a unique
global regular solution for all initial data with finite enstrophy.
Moreover, it has been shown that the attractor is finite dimensional
with dimension not exceeding $(\log_2 R)+\frac12\log_2(\frac{13}4 3)$
[see Eq. (62) in Ref. \cite{CLT007}], and that the evolution of the
shells less than or equal to $K$ determines the evolution of the remaining modes if $K$
is large enough, i.e. larger than the Kolmogorov wavenumber
(defined below).

When $\nu = F = 0$, the model (\ref{eq:sabra_standard_equations}) has 
two quadratic invariants  depending on the values of the  
parameters $a$, $b$, and $c$. The choice $a=1$, $b=-0.5$, and $c=0.5$ guarantees that the 
non-linear evolution (\ref{eq:sabra_standard_NLT}) conserves the total 
energy (hereafter $\sum_n$ denotes the sum over all the shells)
\begin{equation}
\label{eq:total_energy_def}
E = \sum_{n} |u_n|^2 
\end{equation}
and the total helicity $H = \sum_{n} (-)^n k_n |u_n|^2$, as in the three-dimensional Navier-Stokes equations.

After multiplying Eq. (\ref{eq:sabra_standard_equations}) times 
$u_n^*$, adding the complex conjugate, and summing over all the shells 
from $0$ to $M$, one obtains the equation for the time evolution for 
the energy contained in the first  $M$ shells:
\begin{equation}
\label{eq:sabra_energy_evolution}
\dot{E}_{M} = \Pi^E_M - 2 \nu \sum_{n=0}^M k_n^2 |u_n|^2 + 2 \sum_{n=0}^{M} \mathrm{Re}(F_n u_n^*) \, ,
\end{equation}
where 
\begin{align}
	\Pi^E_M = & \, -2 k_M \big[ 2a \mathrm{Im}(u_{M+2} u_{M+1}^* u_{M}^*)
	+ \nonumber \\ & + (a+b) \mathrm{Im}(u_{M+1} u_{M}^* u_{M-1}^*)
	\big]
	\label{eq:sabra_standard_flux}
\end{align}
is the (instantaneous) energy flux through the $M$th shell.

The model given in Eq. (\ref{eq:sabra_standard_equations}) 
spontaneously develops an energy cascade from the large (forced) scales 
to the small ones, with a constant energy flux at steady state. The 
energetics of such a system is given by Eq. 
(\ref{eq:sabra_energy_evolution}) with $M=N-1$ and reads
\begin{equation}
\label{eq:sabra_energetics}
\dot{E} = \epsilon - 2 \nu \Omega\, ,
\end{equation}
where the rate of energy injection $\epsilon=2\sum_n \mathrm{Re}(F_n u_n^*)$
is bounded by $2 |F| \sqrt{E}$ and $2\nu \Omega$ is the rate of
energy dissipation, with
\begin{equation}
\Omega=\sum_n k_n^2 |u_n|^2
\label{eq:omega_def}
\end{equation}
the total enstrophy. The energy flux in 
(\ref{eq:sabra_energetics}) is equal to zero because the nonlinear 
term (\ref{eq:sabra_standard_NLT}) conserves energy.

From Eq. (\ref{eq:sabra_energetics}) at a stationary state and the
Schwarz inequality, it follows that the average of $E$ is $\bar E \le
|f|^2 (k_0 \nu)^{-2}$ , implying the boundedness of the phase space
asymptotically visited by the system.

At a stationary state, realized when $\langle\epsilon\rangle=\langle2
\nu \Omega\rangle$, the energy injected at large scales cascades
towards the small scales with a constant flux $\langle \Pi^E_n\rangle
= - \langle \epsilon\rangle$ for all shells between the forcing one
and the Kolmogorov wave number $k_\eta =
\langle\epsilon\rangle^{1/4} \nu^{-3/4}$, where dissipation becomes
dominant over nonlinear transfers (provided $k_\eta<k_{N-1}$)
\cite{biferale2003shell}.

\subsection{Class of Reversible Shell Models\label{subsec:models_shellrev}}
\begin{figure}[hb!]
	\centering
	\includegraphics[width=0.45\textwidth]{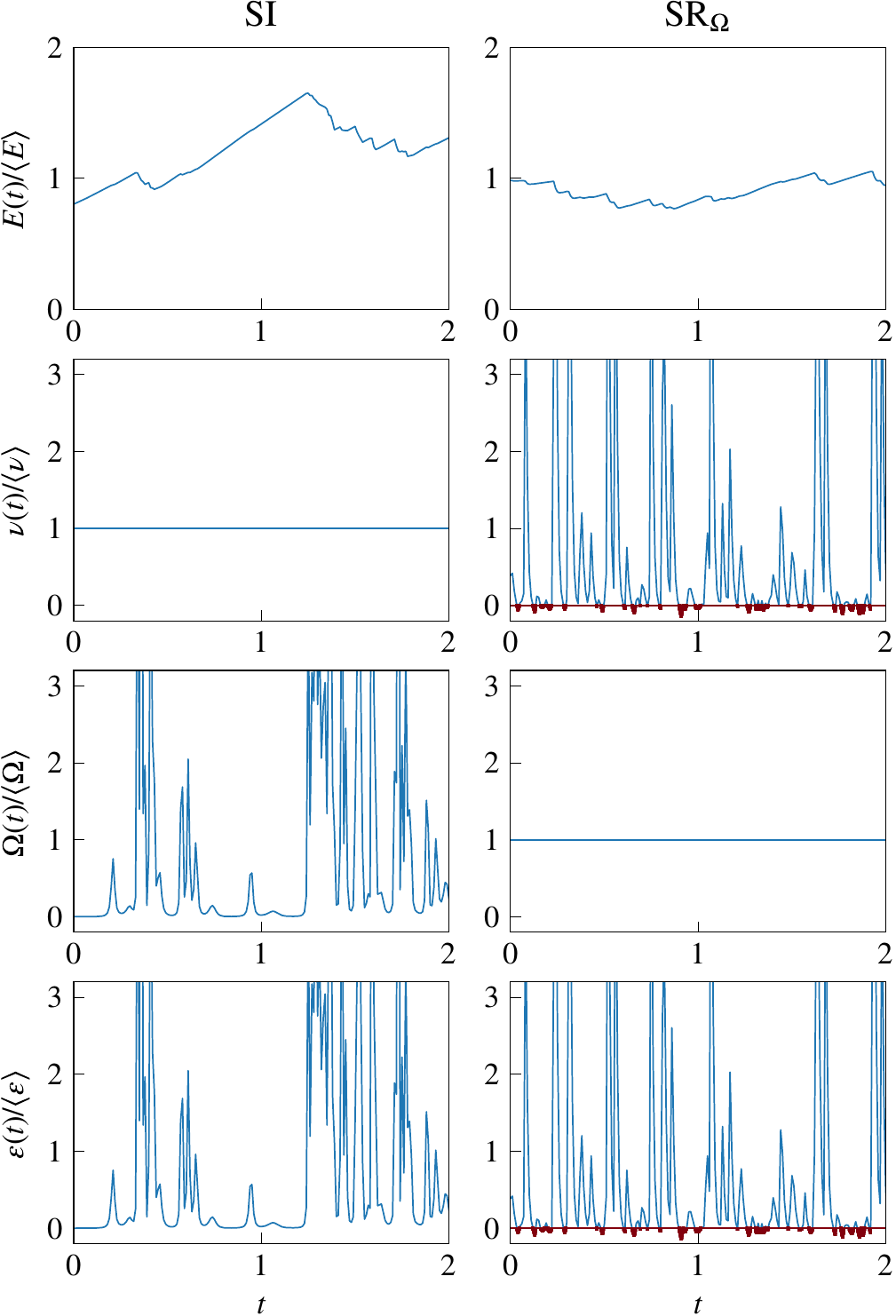}
	\caption{ Side by side comparison between the time
		evolution of several observables in the irreversible shell model
		$\mathrm{SI}$ (left column) and the reversible $\mathrm{SR}_\Omega$
		model (right column). Shown, from top to bottom are the energy $E$, viscosity
		$\nu$ ($\alpha_2$ for $\mathrm{SR}_\Omega$), enstrophy $\Omega$, and
		energy dissipation $\varepsilon = 2 \nu \Omega$ ($2 \alpha_2 \Omega$
		for $\mathrm{SR}_\Omega$). All quantities are normalized by their
		average value. We used $N=15$ shells and $\nu=10^{-5}$,
		corresponding to the energy-cascade regime. Notice that $\alpha_2$
		is not positive definite: The occurrence of negative values,
		highlighted with a thick red line, corresponds to instances in
		which the dissipative terms inject energy into the system.}
	\label{fig:panoramica} 
\end{figure} 

Following the procedure discussed in Sec. \ref{sec:equivalence} (see
also Refs. \cite{gallavotti1996equivalence,Gallavotti1997Dynamical}),
we can define a class of time-reversible models out of
Eq.~(\ref{eq:sabra_standard_equations}) as
\begin{equation}
\label{eq:sabra_reversible_generic_equations}
\dot{u}_n = \mathcal{N}[\{u_n\}] - \alpha_\chi[\{u_n\}]  k_n^2 u_n + F_n \, ,
\end{equation}
where the fluctuating viscosity $\alpha_\chi$  is a function of the 
velocity variables $\{u_n\}$ chosen so as to conserve a 
generic quadratic quantity of the form
\begin{equation}
\label{eq:constraint_generic_invariant}
O_{\chi} \equiv \sum_{n} k_n^\chi |u_n|^2 = \mathrm{const} \,,
\end{equation}
where the continuous parameter $\chi$ weighs differently the wave numbers. 
With this choice, the fluctuating viscosity takes the form
\begin{equation}
\begin{aligned}
\alpha_\chi = & \frac{ \sum_{n} k_n^\chi \mathrm{Re}(u_{n} F^*_n)}{ \sum_n k_n^{\chi+2} |u_n|^2 } \\
& + \frac{ \sum_{n} k_n^{\chi+1} [2a  C_{3,n+1} + b C_{3,n} - (c/2) C_{3,n-1}] }{ \sum_n k_n^{\chi+2} |u_n|^2 }\, ,
\end{aligned}
\label{eq:reversible_shellmodel_nu_expression_generic_invariant}
\end{equation}
where $C_{3,n} \equiv -\mathrm{Im}(u_{n+1} u_n^* u_{n-1}^*)$. Notice that 
the constraint (\ref{eq:constraint_generic_invariant}) implies that also the 
reversible models evolve in a bounded region of the phase space.

For simplicity, we will denote the irreversible shell model 
$\mathrm{SI}$ and the reversible ones as $\mathrm{SR}_\chi$, where 
$\chi$ is the same parameter as in Eq.~(\ref{eq:constraint_generic_invariant}), representing the observable 
kept constant by the time-dependent viscosity. Two limiting cases of 
interest are $\chi=0$ and $\chi=2$, which we will also be indicated as 
$\mathrm{SR}_E$ and $\mathrm{SR}_\Omega$, respectively.

The case $\chi=0$ corresponds to setting the total energy $O_{\chi = 
	0}=E$. Since the energy is conserved in the inviscid limit, the second 
term on the right-hand side of 
(\ref{eq:reversible_shellmodel_nu_expression_generic_invariant}) is 
zero.  Notice also that while the constraint $\chi=0$ is apparently 
applied equally on all wave numbers, in the presence of an energy 
cascade it weighs more the first shells (large scales).

The second case corresponds to setting the enstrophy 
(\ref{eq:omega_def}) $O_{\chi=2}=\Omega$, which, due to the factor 
proportional to the square of the wave number, puts most of the weight 
on the small scales. The limit $\chi=2$ is particularly interesting as 
the energy dissipation rate in the original model 
(\ref{eq:sabra_standard_equations}) $\varepsilon(t)= 2 \nu \Omega$ 
fluctuates by virtue of the fluctuations of $\Omega$, while in the 
reversible model (\ref{eq:sabra_reversible_generic_equations})  
$\varepsilon(t)= 2 \alpha_2 \Omega$ fluctuates with the viscosity 
$\alpha_2$. This phenomenology is clear from  Fig. \ref{fig:panoramica}, 
where we present an overview of the dynamics of the standard 
$\mathrm{SI}$ model and the $\mathrm{SR}_\Omega$ ($\chi=2$) model. The 
figure shows the time evolution of some observables of interest such as 
$E$, $\Omega$, $\nu$, and $\varepsilon$, in a situation of energy cascade 
for both systems. These results will be analyzed in more detail in 
the next section.

We remark that some preliminary study of the model with $\chi=2$ was 
presented in Ref. \cite{depietro}, where the aim was not 
that of studying the equivalence in the sense specified in Sec. 
\ref{sec:equivalence}.

\section{Numerical Simulations}\label{sec:numerical}
\subsection{Setup of the Numerical Simulations}

We first integrate  the irreversible ($\mathrm{SI}$) model 
(\ref{eq:sabra_standard_equations})  for as long as  needed to achieve 
stationarity. The resulting average energy spectrum $E_n |_{\nu} \equiv 
\langle |u_n|^2 \rangle |_{\nu}$ is then used to set the initial 
condition for the reversible model 
(\ref{eq:sabra_reversible_generic_equations}) as 
\begin{equation}
u_{n} (t=0)|_{\chi}
\equiv \sqrt{ E_n |_{\nu}}\, e^{i \xi_n} \,,
\label{eq:initial}
\end{equation}
where $\xi_n$ are random phases. The initial condition 
(\ref{eq:initial}) guarantees that the reversible dynamics starts with 
initial values for the considered global quadratic observable $O_\chi$  
[see Eq.~(\ref{eq:constraint_generic_invariant})], such as $E$ or 
$\Omega$, equal to the expectation value obtained with the irreversible 
model. 

Simulations of $\mathrm{SI}$ are performed fixing by the number of shells, 
holding constant the large scale forcing and varying the viscosity 
$\nu$, which here plays the role of the inverse of the Reynolds number  
$R = 1/\nu$. A change in the chosen value of the viscosity $\nu$ 
is reflected in an initial configuration for the reversible models with 
different values of the conserved quantity $O_\chi$. 

The corresponding reversible model is then integrated with the same 
number of shells $N$ and forcing $F_n$ as the irreversible case. As for 
the forcing,  we have chosen a constant (hence time reversible) 
forcing acting on the first shell only, i.e., $F_n = \delta_{n,0} \, |F| 
e^{i \gamma}$ with $|F| = 1$ and $\gamma$ a randomly chosen phase. 

A statistical ensemble of ten dynamical evolutions was obtained by
varying the phases $\xi_n$ of the initial condition. All data
presented hereafter are averages on this ensemble and the errors are
estimated as the standard error on the mean. The characteristic time
of the large scales is estimated as $T_L \sim E/\langle \varepsilon
\rangle$, which is $O(1)$ in our simulations.  The total integration
time (cumulated over all the simulations in the ensemble) ranges
between $\sim 10^5 T_L$ for the smallest Reynolds number and $\sim 10^4
T_L$ for the largest.

The integration scheme was a modified fourth-order Runge-Kutta scheme with 
explicit integration of the linear part (see the Appendix 
for details). For both the irreversible and 
reversible models, the (fixed) integration time step was guaranteed to 
be $\delta t \le \tau_{\mathrm{min}}/50$, with $\tau_{\mathrm{min}} = 
\min_n (\langle |u_n| \rangle k_n)^{-1}$ the fastest time scale 
of the dynamics. The number of shells in the system was $N = 20$, 
unless otherwise specified.

\subsection{Test of the equivalence in the reversible model conserving the total enstrophy}

We start by discussing the reversible model $\mathrm{SR}_\Omega$
obtained by imposing the conservation of enstrophy. Reversible models
conserving other quadratic quantities (in particular, $\mathrm{SR}_E$
which conserves energy) will be discussed in the next section.

\begin{figure}[b!]
	\centering
	\includegraphics[width=0.45\textwidth]{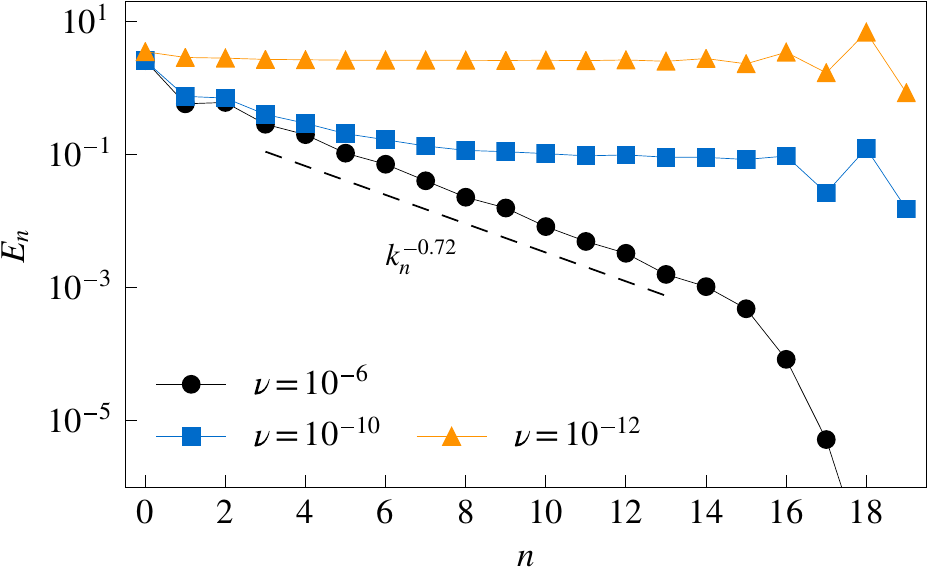}
	\caption{ Phenomenology of the irreversible shell
		model.  The energy spectrum $E_n=\langle |u_n|^2\rangle$ is shown for three
		values of the viscosity $\nu$ with $N=20$ shells. With $\nu=10^{-6}$
		the characteristic spectrum of the energy cascade $E_n\sim
		k_n^{-0.72}$ (see the dashed line) appears. For $\nu=10^{-12}$ the
		energy spectrum is roughly at equipartition, namely, the regime of
		quasiequilibrium (see the text). For $\nu=10^{-10}$ a mixed
		behavior is observed.  }
	\label{fig:SI_spectra}
\end{figure}

\paragraph*{Phenomenology of the  irreversible model:} First it is 
useful to illustrate briefly the phenomenology of the irreversible
model with a fixed number of shells with increasing Reynolds number,
viz. decreasing the viscosity value $\nu$. In
Fig.~\ref{fig:SI_spectra} we show the energy spectrum obtained for
three values of $\nu$. When the viscosity is small enough but such
that $k_\eta \ll k_{N-1}$ (i.e., when the dissipative scale is well
resolved), the irreversible shell model develops an energy cascade,
from large to small scales, with a characteristic Kolmogorov-like
scaling $E_n=\langle |u_n|^2\rangle \propto k_n^{-2/3}$ plus
intermittency corrections \cite{biferale2003shell}. This
energy-cascade regime is evident for $\nu=10^{-6}$ in
Fig.~\ref{fig:SI_spectra}. When the Reynolds number is very large,
viz. the viscosity is so small that $k_\eta \gg k_{N-1}$, a
new stationary regime sets in. In the following such a regime will be
referred to as quasiequilibrium as it is characterized by the
energy being essentially equipartitioned (though in nonequilibrium
conditions) among the shells (see the case $\nu=10^{-12}$ in
Fig.~\ref{fig:SI_spectra}) and by an average energy flux constant
over the shells and typically much smaller than its fluctuations (not
shown). The transition between the energy-cascade and
quasiequilibrium regimes is characterized by energy spectra with
intermediate characteristics (see the case $\nu=10^{-10}$ in
Fig.~\ref{fig:SI_spectra}). Strictly speaking, the dynamical
equivalence discussed in Sec.~\ref{sec:equivalence_theo} is expected
to hold in the quasiequilibrium regimes, when the Reynolds number
$R\sim \nu^{-1}$ is large enough and $N$ is fixed. Note that in
\cite{gallavotti2014equivalence} the validity of the equivalence has
been confirmed exactly in such quasiequilibrium conditions.

\begin{figure}[h!]
	\centering
	\includegraphics[width=0.45\textwidth]{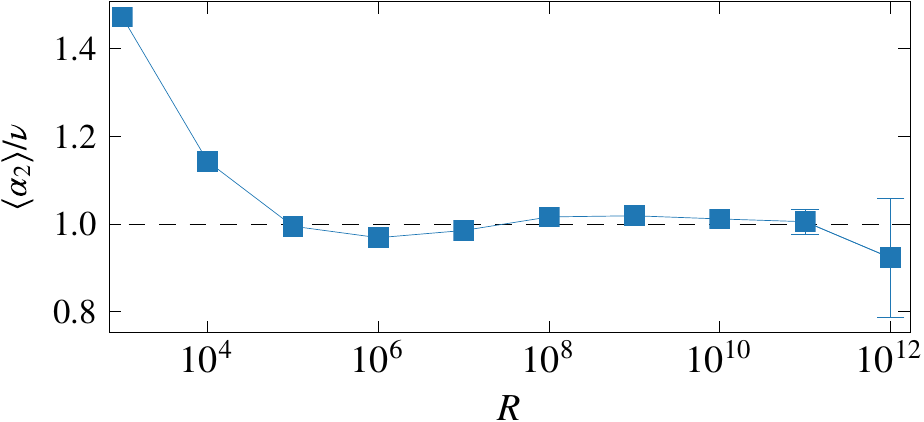}
	\caption{ Mean values of $\alpha_2 / \nu$ for simulations of the
		$\mathrm{SR}_\Omega$ model at different Reynolds number $R$ with
		$N=20$ shells. The $R$ dependence of the $\mathrm{SR}_\Omega$ model
		is intended in the sense that it is initialized with an initial
		enstrophy $\tilde{\Omega}$ equal to $\langle \Omega \rangle$
		measured in a run of the $\mathrm{SI}$ model with (fixed) viscosity
		$\nu = R^{-1}$. The large error bars reported for high $R$ can be
		ascribed to the limited statistics, due to the high cost of the
		numerical integration in that range of parameters.  }
	\label{fig:convergenza_nu} 
\end{figure}

\paragraph*{Test of the Equivalence Conjecture:} As a preliminary test 
of the equivalence, we first verified the validity of
Eq.~(\ref{e1.4a}), i.e., we checked whether the average value
$\langle \alpha_2 \rangle$ measured in the $\mathrm{SR}_\Omega$ model
simulations converges to the values of the viscosity $\nu$ of the
corresponding irreversible model. In Fig.~\ref{fig:convergenza_nu}
we show the ratio $\langle \alpha_2\rangle/\nu$ at varying $R$ ($R =
\nu^{-1}$, where $\nu$ is the viscosity of the $\mathrm{SI}$
model). As one can see, the ratio is approximately equal to $1$ for $R > 10^{-5}$ and
unity is approached more closely with increasing $R$, apart from
the highest $R$, where computational constraints on the integration
time lead to poorer convergence of the statistics and thus larger
statistical errors.

\begin{figure}[t!]
	\centering
	\includegraphics[width=0.45\textwidth]{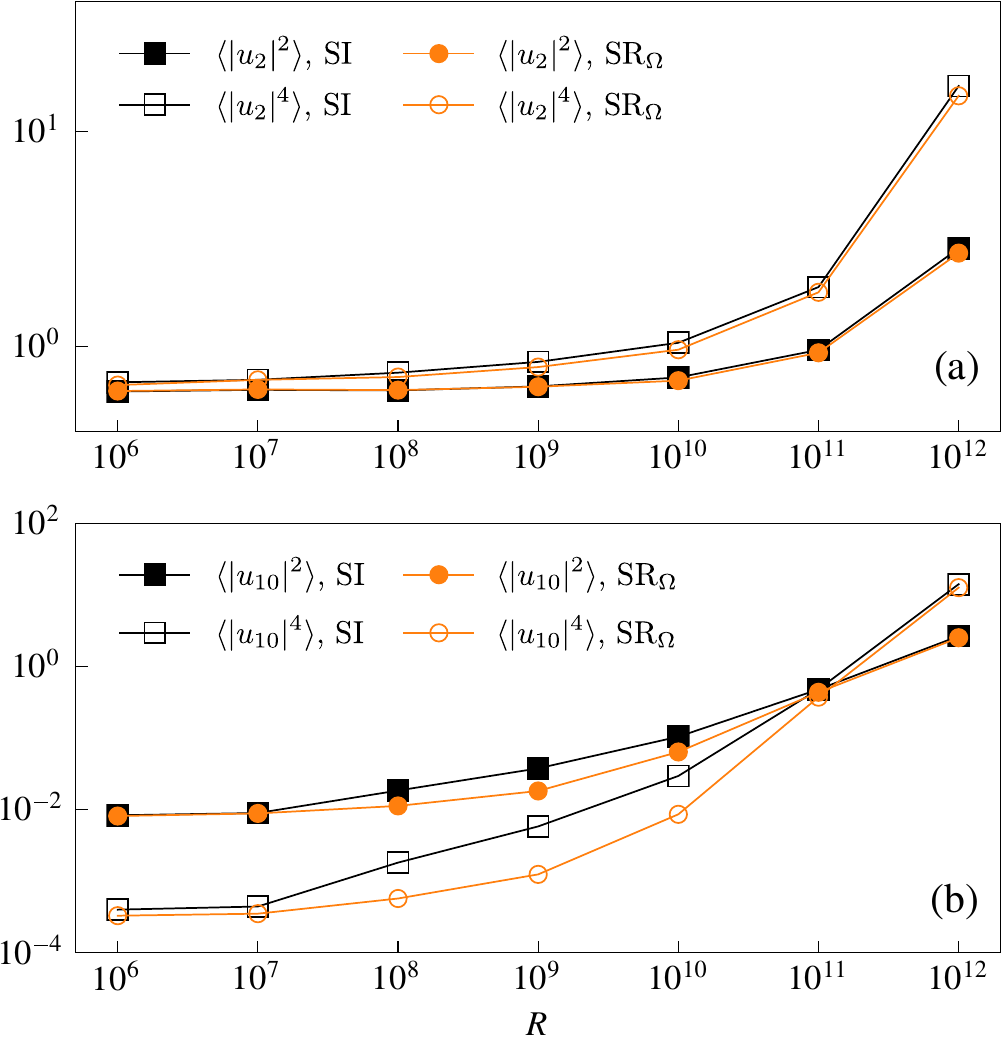}
	\caption{ Test of the equivalence for the
		$\mathrm{SR}_\Omega$ model. The second moment (closed symbols) and
		fourth moment (open symbols) of a velocity field component
		pertaining to the (a) large scales $n=2$ and (b) small scale $n=10$ are shown
		as functions of the Reynolds number $R$ for both the $\mathrm{SI}$
		and $\mathrm{SR}_\Omega$ models with $N=20$ shells. The $R$
		dependence of the $\mathrm{SR}_\Omega$ model is intended in the
		sense that it is initialized with an initial enstrophy
		$\tilde{\Omega}$ equal to $\langle \Omega \rangle$ measured in a run
		of the $\mathrm{SI}$ model with (fixed) viscosity $\nu =
		R^{-1}$. Errors are smaller than or of the order of the symbol size.
	}
	\label{fig:momenti_large}
\end{figure}
As discussed in Sec. \ref{sec:equiv_conj_statement}, the validity of
Eq.~(\ref{e1.4a}) is a prerequisite for the equivalence conjecture.
Then we tested the conjecture (\ref{eq:conjecture1}) at varying
values of $R$ using as the observable $\Phi$ the second and fourth
moments of $|u_n|$ for a small wave number [shell $n=2$; see
Fig.~\ref{fig:momenti_large}(a)]. These moments are effectively large-scale observables and the equivalence conjecture is expected to hold
for them for high values of $R$. We also measured the same moments at
a larger wave number [shell $n=10$; see Fig.~\ref{fig:momenti_large}(b)]
where, in principle, the validity of the equivalence should not be
taken for granted. At large scales [Fig.~\ref{fig:momenti_large}(a)] the
data points of the $\mathrm{SR}_\Omega$ perfectly agree with the
values of the $\mathrm{SI}$ model at all the $R$ considered. At
smaller scales [Fig. \ref{fig:momenti_large}(b)] we observe good
agreement with $\mathrm{SI}$ at high and relatively small $R$,
i.e. in both the quasiequilibrium and energy-cascade regime,
while deviations are present at intermediate values of $R$.

\begin{figure*}[t!]
	\centering
	\includegraphics[width=0.95\textwidth]{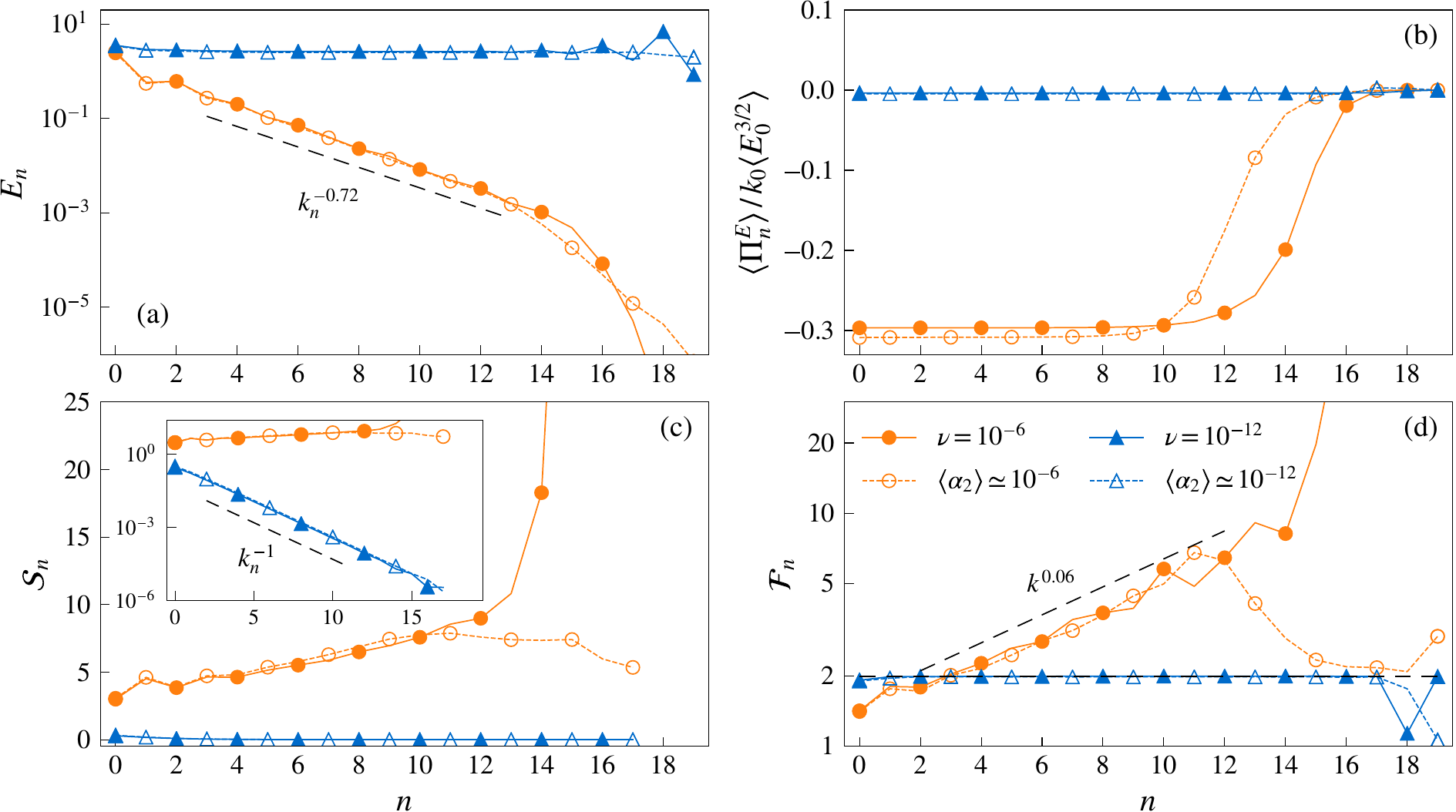}
	\caption{ Comparison of several spectral observables
		between the $\mathrm{SI}$ and the $\mathrm{SR}_\Omega$ models in
		both situations of energy cascade [$\nu=10^{-6}$
		(\textcolor{PlotOrange}{\large{$\bullet$}}) and $\langle \alpha_2 \rangle
		\simeq 10^{6}$ (\textcolor{PlotOrange}{\large{$\circ$}})] and
		quasiequilibrium [$\nu=10^{-12}$
		(\textcolor{PlotBlue}{\small{$\blacktriangle$}}) and $\langle \alpha_2
		\rangle \simeq 10^{12}$
		(\textcolor{PlotBlue}{\scriptsize{$\triangle$}})]:
		(a) energy spectra, (b) energy flux (\ref{eq:sabra_standard_flux}),
		(c) skewness $\mathcal{S}_n$ (\ref{eq:skewness}) (the inset shows the
		same plot with the logarithmic $y$ axis), and (d) flatness $\mathcal{F}_n$
		(\ref{eq:flatness}). Errors are the order of or smaller than the
		symbol size. The dashed line labeled $k_n^{-0.72}$ in (a) represents
		the scaling behavior in the manner of Kolmogorov plus intermittency
		correction. The dashed line labeled $k_n^{-1}$ in the inset of 
		(c) represents a dimensional prediction valid at quasiequilibrium;
		indeed, since $\langle|u_n|\rangle$ and $\langle \Pi_n^E \rangle$ do
		not depend on the wave number $k_n$, at least in a certain range of
		scales, as shown in (a) and (b) respectively, one has that
		$\mathcal{S}_n \sim k_n^{-1}$ [Eq.~(\ref{eq:skewness})]. The dashed line
		labeled $k_n^{0.06}$ in (d) shows a best fit of the curves in
		the cascade regime.  Finally, the horizontal dashed line in 
		(d) displays the value $\mathcal{F}_n = 2$, which is expected for
		complex Gaussian variables. In these figures, and in some of the
		following ones, to ease the identification of the various curves and
		avoid the superposition of different symbols, not all data points
		have been marked by a symbol.  }	\label{fig:Espectra_separated_I_RO} 
\end{figure*}

In order to understand better the above findings, in
Fig.~\ref{fig:Espectra_separated_I_RO}(a) we compare the energy spectra
of the $\mathrm{SI}$ and $\mathrm{SR}_\Omega$ in the two regimes of
energy cascade and quasiequilibrium. Consistently with
Fig.~\ref{fig:momenti_large}, a very good equivalence between the
reversible and irreversible models is observed in both regimes, at
least at large enough scales. At small scales deviations can be seen
in both regimes.

In the cascade regime, the main differences appear for $k_n>k_\eta$. It 
should be noticed that $k_\eta > k_{10}$, which explains the agreement 
observed in  Fig.~\ref{fig:momenti_large}(a). Clearly, choosing a 
wavenumber $k_n>k_\eta$ does lead in general to good agreement. It 
is worth noticing that for $k_n>k_\eta$ the energy spectrum has a 
scaling law close to $E_n \sim k_n^{-2}$, which could be due to a local 
equipartition of the enstrophy (which is mostly localized around these 
scales).\footnote{Further simulations (not shown) performed at the best 
	of our computational possibilities (resolutions up to $N=45$ and values 
	of $\Omega$ up to $\sim 10^{12}$) did not show a clear trend towards 
	such an equipartition and we consider the question still open.} 

In the quasiequilibrium regime, we can notice that the
$\mathrm{SR}_\Omega$ model shows a more regular spectrum at small
scales (near the boundary $k_{N-1}$), with respect to the
$\mathrm{SI}$ model.  We should remark that these oscillations in the
$\mathrm{SI}$ model remain confined to the last three or four shells, as
confirmed by simulations with a larger number of shells (not
shown). Our interpretation is that they are simply due to the
constraint imposed by the fixed ultraviolet cutoff, which becomes
important when the scales affected are not efficiently damped by
viscosity. The choice $\Omega = \mathrm{const}$ imposes a constraint on the
amount of energy present at scales around $k \sim \sqrt{\Omega/E}$,
suppressing such oscillations coming from the spectral truncation.

We also compared other quantities in the two models at varying 
Reynolds number. In particular, we studied the average energy flux
(\ref{eq:sabra_standard_flux})
[Fig.~\ref{fig:Espectra_separated_I_RO}(b)]; the skewness
[Fig.~\ref{fig:Espectra_separated_I_RO}(c)] defined as
\begin{equation}
\label{eq:skewness}
\mathcal{S}_n = \frac{ \langle \Pi^E_n \rangle }{k_n \langle |u_n| \rangle 
	\langle |u_{n+1}| \rangle \langle |u_{n+2}| \rangle} \, ,
\end{equation}
where we use products $|u_n| |u_{n+1}||u_{n+2}|$ in place of $|u_n|^3$ 
to get rid of spurious oscillations due to the phase symmetry between 
three adjacent shells (see \cite{Lvov_1998_improved_shellmodels} for 
details); and the 
flatness [Fig.~\ref{fig:Espectra_separated_I_RO}(d)]
\begin{equation}
\label{eq:flatness}
\mathcal{F}_n = \frac{\langle |u_n|^4 \rangle}{\langle |u_n|^2 \rangle^2} \, .
\end{equation}
In the cascade regime, the equivalence holds only within the inertial
range of scales, which is slightly shorter in $\mathrm{SR}_\Omega$
compared to $\mathrm{SI}$; indeed, as clear from
Fig.~\ref{fig:Espectra_separated_I_RO}(b), the flux for
the $\mathrm{SR}_\Omega$ model stops being constant at slightly smaller
wave numbers than in the $\mathrm{SI}$ model. We observe remarkable
agreement also for very delicate properties such as the intermittent
corrections to the scaling exponents as clear from both the energy
spectrum [Fig.~\ref{fig:Espectra_separated_I_RO}(a)] and high-order
quantities such as $\mathcal{S}_n$ and $\mathcal{F}_n$
[Figs.~\ref{fig:Espectra_separated_I_RO}(c) and~\ref{fig:Espectra_separated_I_RO}(d)]. A previous study
confirmed this equivalence also on higher order structure functions
$\langle |u_n|^q\rangle$, up to order $q=9$ \cite{depietro}. These
results offer further confirmation of the extreme robustness of the
energy-cascade mechanism with respect to the particular method used to
remove energy at small scales, thus reinforcing the
validity of the dynamical equivalence.

Also in the quasiequilibrium regime (i.e., for the simulation
corresponding to $\nu=10^{-12}$) a very good equivalence is observed
for all the quantities. In particular, we notice that in the
quasiequilibrium regime the statistics tends to become Gaussian with
$\mathcal{S}_n\to 0$ and $\mathcal{F}_n\approx 2$ (which is the result
expected for Gaussian statistics, taking into account the fact that
$u_n$ is complex).

Between these two regimes, for intermediate values of the viscosity,
deviations are well evident [as already clear from
Fig.~\ref{fig:momenti_large}(b)].

Summarizing, the equivalence conjecture is well verified in the 
quasiequilibrium regime, where it is expected to hold, at almost all 
scales excluding those very close to the ultraviolet cutoff. 
Remarkably, the equivalence holds, even for very delicate quantities, 
also in the energy-cascade regime at scales $k_n \lesssim k_\eta$. We 
notice that the equivalence in the latter case may have a different 
nature from that of the former. In particular, when the energy cascade 
is at play, the matching of the statistics of the various observables 
within the inertial range may be due to the robustness of the inertial 
range physics with respect to the energy removing mechanisms, i.e., due 
to the dissipative anomaly.

\subsection{Test of the Equivalence in Reversible Models Conserving Different Quantities}

Here we  discuss the equivalence in the reversible models 
(\ref{eq:sabra_reversible_generic_equations}) with varying the parameter 
$\chi$ in (\ref{eq:constraint_generic_invariant}), i.e., with varying the 
particular quadratic quantity conserved by the time-dependent 
viscosity. 

\begin{figure}[t!]
	\centering
	\includegraphics[width=0.45\textwidth]{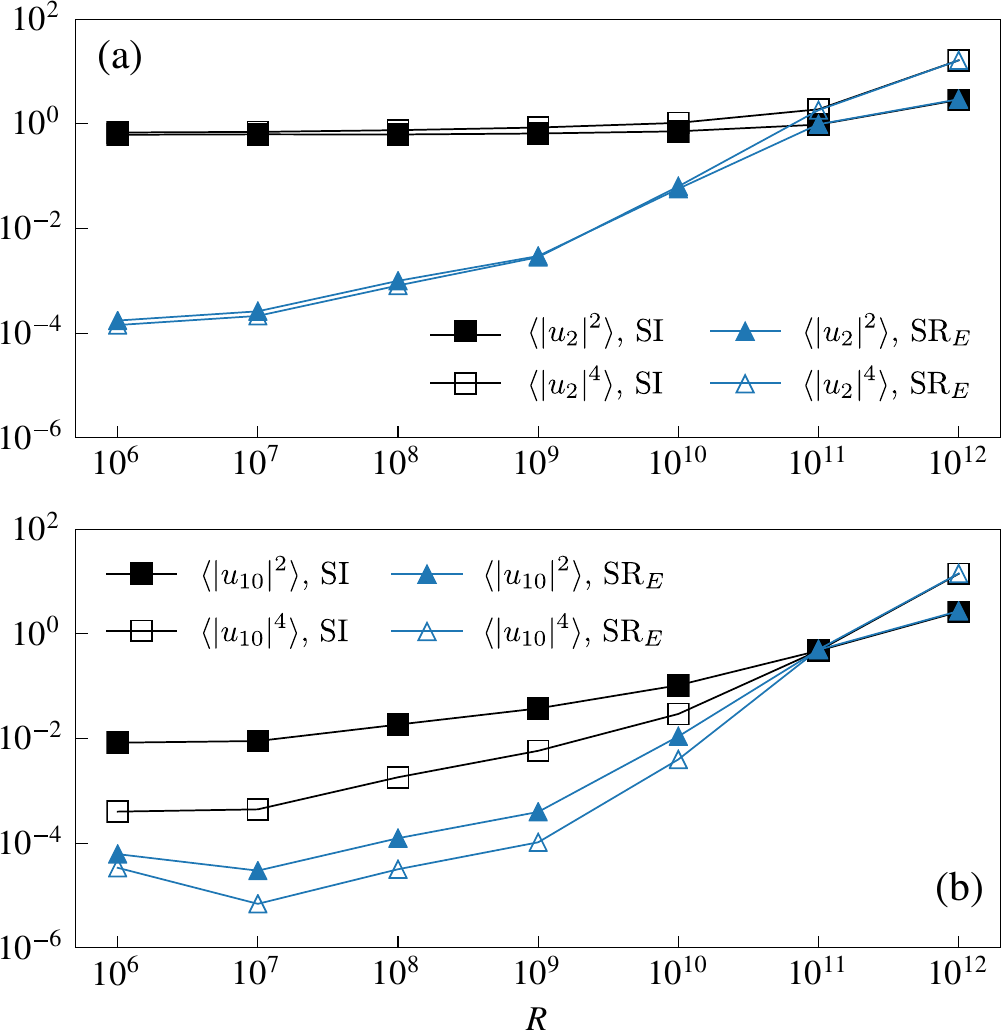}
	\caption{ Test of the equivalence for the
		$\mathrm{SR}_E$ model.  The second moment (closed symbols) and fourth moment (open
		symbols) of a velocity component at (a) large scales $n=2$ and (b) small
		scales $n=10$ as functions of the effective Reynolds number $R$
		for the $\mathrm{SI}$ and $\mathrm{SR}_E$ models with $N=20$ shells.
		The $R$ dependence of the $\mathrm{SR}_E$ model is intended in the
		sense that it is initialized with an initial energy $\tilde{E}$
		equal to $\langle E \rangle$ measured in a run of the $\mathrm{SI}$
		model with (fixed) viscosity $\nu = R^{-1}$. Errors are smaller than
		or of the order of the symbol size.} \label{fig:momenti_SME}
\end{figure} 
\begin{figure}[t!]
	\centering	
	\includegraphics[width=0.45\textwidth]{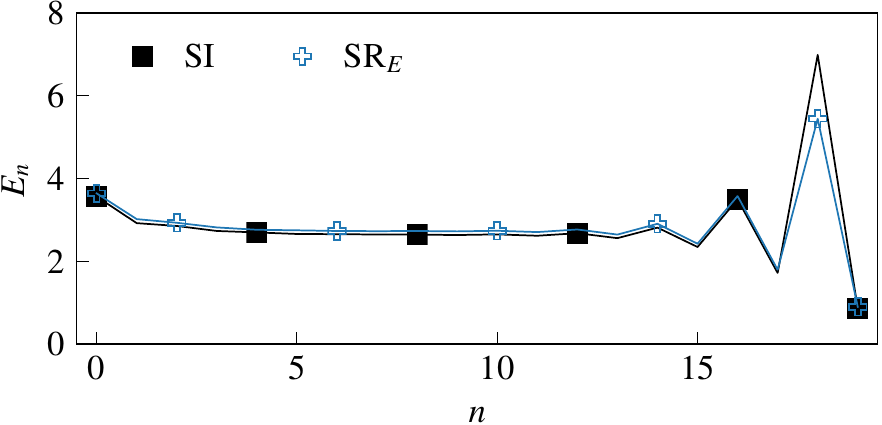}
	\caption{ Energy spectra $E_n$ of the $\mathrm{SI}$ and
		$\mathrm{SR}_E$ models in the regime of quasiequilibrium ($N=20$ and
		$\nu = 10^{-12}$). Error bars are smaller than or of the order of
		the symbol size.  }\label{fig:spettri_SME} 
\end{figure} 
\begin{figure}[t!]
	\centering
	\includegraphics[width=0.45\textwidth]{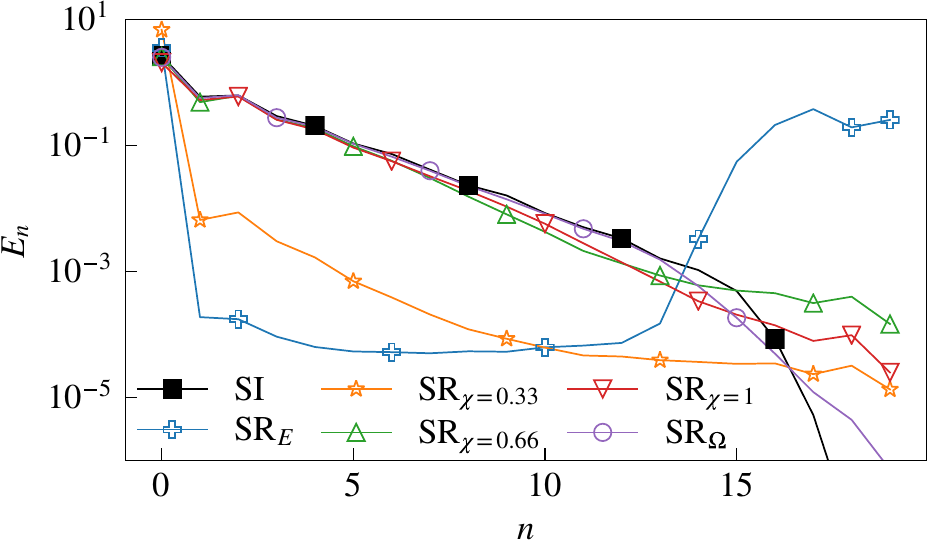}
	\caption{ Energy spectra $E_n$ for several reversible
		models, compared with the irreversible one
		({\scriptsize{$\blacksquare$}}) (with $N=20$ and $\nu=10^{-6}$). All
		the reversible model simulations are initialized with the same
		distribution of initial energy in the range $0 \le n < 15$, but the
		models conserve different invariants $O_\chi$ [see Eq.
		(\ref{eq:constraint_generic_invariant})]. Error bars are smaller
		than or of the order of the symbol size.}  \label{fig:espectra_varying_alpha} 
\end{figure} 
\begin{figure*}[hbt!]
	\centering
	\includegraphics[width=0.8\textwidth]{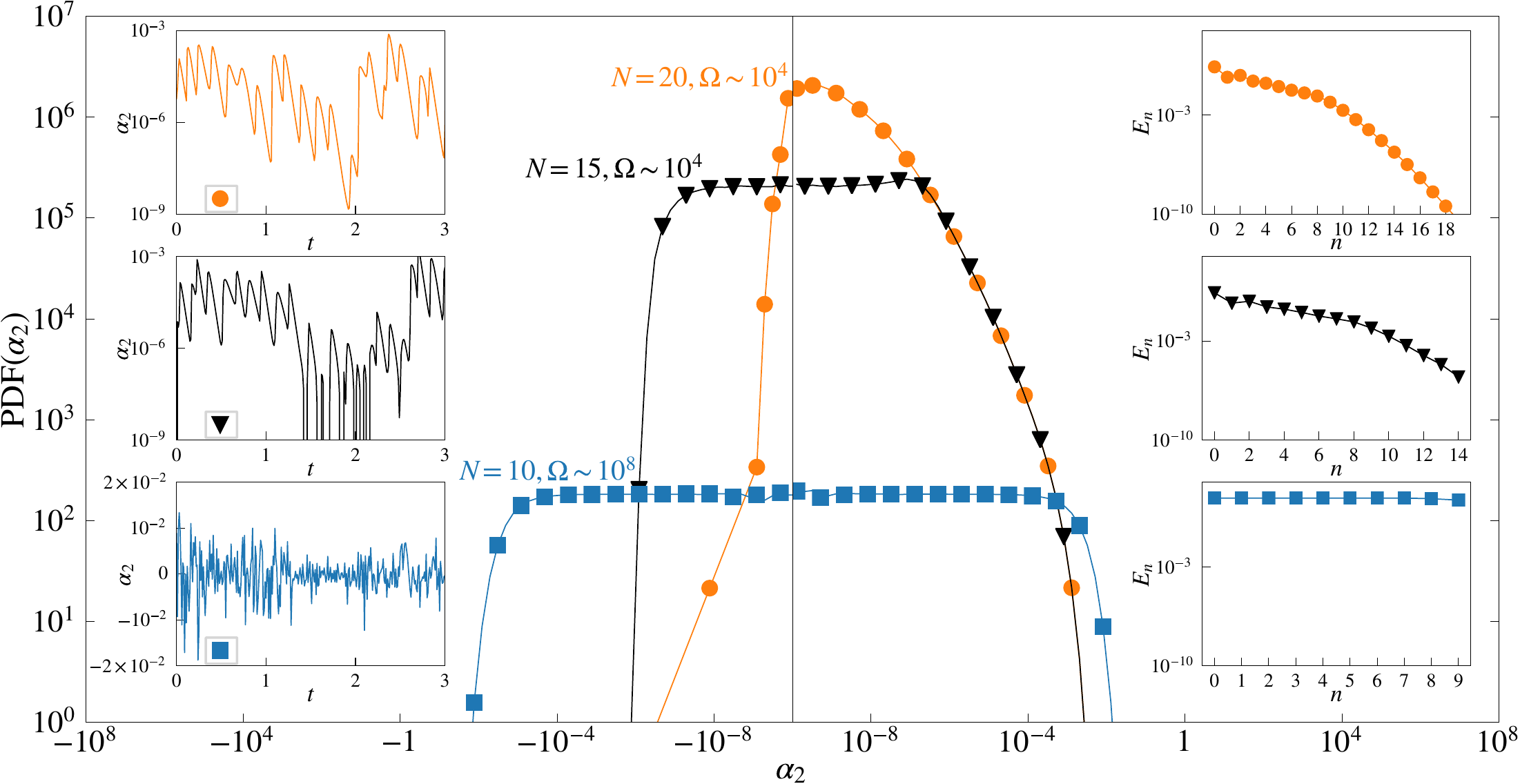}	
	\caption{ Probability density function of the
		time-dependent viscosity $\alpha_2$ for the $\mathrm{SR}_{\Omega}$ model
		in three different cases: a situation of energy cascade [$N=20$ and
		$\Omega\sim10^{4}$ (\textcolor{PlotOrange}{\large{$\bullet$}})], a
		situation of quasiequilibrium [$N=10$ and
		$\Omega\sim10^{8}$ (\textcolor{PlotBlue}{\scriptsize{$\blacksquare$}})],
		and a case in between [$N=15$ and
		$\Omega\sim10^{4}$ ($\blacktriangledown$)]. The insets on the left
		show the corresponding typical time evolutions of $\alpha_2$. The insets
		on the right show the corresponding energy spectra.  }
	\label{fig:nu_O} 
\end{figure*} 

We start from Fig.~\ref{fig:momenti_SME}, which, analogously to
Fig.~\ref{fig:momenti_large}, shows the $R$ dependence of the second and
fourth moments of $|u_n|$ for $n=2$ [Fig.~\ref{fig:momenti_SME}(a)] and $n=10$ [Fig.~\ref{fig:momenti_SME}(b)] for the
$\mathrm{SR}_E$ model, i.e., when the reversible model is obtained by
imposing the conservation of energy. Unlike the $\mathrm{SR}_\Omega$
model shown in Fig.~\ref{fig:momenti_large}, we can see that agreement
between the moments of the $\mathrm{SI}$ and $\mathrm{SR}_E$ models is
realized only in the quasiequilibrium regime. This is further
confirmed in Fig.~\ref{fig:spettri_SME}, where we compare the energy
spectra of the $\mathrm{SI}$ and $\mathrm{SR}_E$ models in this
regime. As clear from the figure, for the $\mathrm{SR}_E$ model the  
agreement of the spectra extends even close to the ultraviolet cutoff
[compare with Fig.~\ref{fig:Espectra_separated_I_RO}(a)]. This is
possibly due to the fact that the constraint of constant energy is
less stringent for the large wave numbers compared
with the constant enstrophy constraint.

In order to understand the large differences between the
$\mathrm{SR}_E$ and $\mathrm{SI}$ models out of the quasiequilibrium
regime, we now fix $\nu$ such that the $\mathrm{SI}$ model is in the
energy cascade-regime. In Fig.~\ref{fig:espectra_varying_alpha}(a) we
show the spectra obtained for different reversible models, all
initialized with the same initial condition, conserving quadratic
quantities $O_\chi$ indexed by different values of $\chi$ as from
Eq.~(\ref{eq:constraint_generic_invariant}) (we recall that
$\mathrm{SR}_E$ corresponds to the case $\chi=0$). We see that there
is a clear trend of increasingly better equivalence with increasing
$\chi$, i.e., when the constraint weights more and more the small
scales. In particular, when the reversible model conserves $O_\chi$
with low values of $\chi$, it suffers from the lack of a stable energy-cascade solution, with the effective confinement of the dynamics on
the shell $n=0$. When the value $\chi$ is large, on the contrary,
$O_\chi$ is significantly dependent on the small scales of the system,
meaning that the request $O_\chi = \mathrm{const}$ actually imposes a
constraint on the amount of energy needed in the small scales,
favoring the presence of a stable energy-cascade mechanism. The
threshold between the two cases lies around $\chi = 2/3$. Even if we
did not pursue a systematic test, here is a simple argument for why
the value $\chi = 2/3$ should be a good candidate for the threshold:
For that value both the constant energy flux solution and the $O_\chi$
equipartition solution have the same spectral scaling $E_n \sim
k_n^{-2/3}$. For $\chi > 2/3$ the constant energy flux solution has a
less steep energy spectrum and it is likely dominant in the dynamics,
and vice versa. Thus, given the same initial conditions for
the velocity field, the  $\mathrm{SR}_\Omega$ model and the other SR
models with $\chi > 2/3$ are always able to reach a chaotic stationary
state with an energy cascade like the $\mathrm{SI}$ model.

Instead, in the same range of viscosities, the SR$_E$ model and the
other SR models with $\chi<2/3$ get locked in a fixed point in phase
space, where all the energy of the system is localized in the $n=0$
shell and $\alpha_0 \sim 1$.

The presence of an attractive fixed point in a highly dimensional 
phase space unavoidably makes the statistical properties strongly 
sensitive to the extension in time of the dynamical evolution and to 
the total number of degrees of freedom. For example, we found that the 
results published in \cite{biferale1998time} were affected by the 
limited extension of the time integration  and that by averaging more, 
as it is possible with the nowadays computational power, the long-time 
asymptotic dynamics is always dominated by the fixed point at small 
shell numbers.

Although we did not perform systematic tests, on the basis of the 
previous observations and  Fig.~\ref{fig:spettri_SME}, it is 
reasonable to expect that for any $\chi$ the equivalence should hold in the 
quasiequilibrium regime.

Specifically, for the  $\mathrm{SR}_E$ model , it is worth remarking
that imposing the conservation of energy constrains the energy
dissipation to be identical to the energy input at any instant. This
is at odds with the phenomenology of the cascade where such a balance
is obtained only on average. On the other hand, setting $\Omega=\mathrm{const}$
does not introduce such stringent conditions on the instantaneous
energy budget. More importantly, while the energy input varies on the
(slow) timescale typical of the large scales, the energy dissipation
has a fast evolution. Thus, the  SR$_E$ model imposes a very severe
dynamical constraint requiring the two quantities to be identical at
each time.  This constraint is less stringent in quasiequilibrium
conditions, where energy is essentially in equipartition among the
shells. Indeed, in such a regime, also the  SR$_E$ model becomes equivalent
to the $\mathrm{SI}$ model as clear from Fig.~\ref{fig:spettri_SME}.

\subsection{Analysis of the Time-Dependent Viscosity in the reversible model with enstrophy conservation}

In this section we study the statistics of the time-dependent
viscosity $\alpha_2$ in the $\mathrm{SR}_\Omega$ model. We have
already shown that $\langle \alpha_2\rangle\approx \nu$
(Fig.~\ref{fig:convergenza_nu}), as required for the validity of the
equivalence. However, the temporal fluctuations of $\alpha_2$ are
nontrivial: As shown in Fig.~\ref{fig:panoramica}, $\alpha_2$ can
become negative (i.e., the viscous forces can inject energy instead of
removing it), which is the signature of the dynamical reversibility.
In this section, though this is not directly linked with testing the
equivalence conjecture, we explore how the statistics of this sign
variation depends on the Reynolds number.

In Fig.~\ref{fig:nu_O} we summarize the behavior of the 
time-dependent viscosity $\alpha_2$ in different regimes: from 
quasiequilibrium to energy cascade (as qualified by the behavior of 
the spectra shown on the right panels). On the left panels we show the 
time evolution of $\alpha_2$ in a typical run of the model, in the 
central panel the measured probability density functions (PDFs) of the values of $\alpha_2$, and on the 
right column the energy spectrum of the corresponding simulation. All 
data refer to the $\mathrm{SR}_\Omega$ model.

In the quasiequilibrium regime, the viscosity $\alpha_2$ tends to
have a PDF symmetric around the zero, becoming more and more skewed
towards positive values as the cascade regime becomes dominant in the
dynamics.  A similar behavior of the PDF of the time-dependent
viscosity of the reversible model as a function of the Reynolds number
was found in \cite{gallavotti2014equivalence}.  In the limit of an
extremely well resolved system ($N \rightarrow \infty$, with finite
Reynolds number, i.e., in the energy-cascade regime with well resolved
dissipative range), the probability to observe negative values
($\alpha_2<0$) within the observation time becomes extremely
small. This observation shows once again the different nature of the
equivalence in the quasiequilibrium regime (corresponding to taking
the limit $R\to \infty$ with $N$ fixed, eventually very large) and the
cascade one (corresponding to taking the limit $N\to \infty$ with $R$
fixed and very large).

\section{Conclusions}
\label{sec:conclusions}

Summarizing, in this paper we have scrutinized the validity of the 
equivalence of ensembles for nonequilibrium statistical mechanical 
systems conjectured for fluid flows in 
\cite{gallavotti1996equivalence,Gallavotti1997Dynamical}. In 
particular, we tested the conjecture  within the framework of  the  
shell models for turbulence featuring a multiscale nonlinear dynamics.

In these systems, the issue of nonequilibrium ensemble equivalence
translates into the quest for equivalence of the macroscopic dynamics
between systems with different modelizations of the viscous
forces. 
The standard choice is to use a constant viscosity, which leads to the introduction, in the evolution equations, of a term that is responsible for breaking the time-reversal symmetry of the equations of motion. The same happens if one introduces instead hyperdiffusive operators, such that the viscosity is effectively larger when smaller spatial scales are considered. However, given the
reversibility of the microscopic dynamics, it is natural to speculate
that a macroscopic description preserving such a fundamental symmetry
should be possible.

Models exhibiting a time-reversal symmetry  can be realized by using 
a time-dependent viscosity designed to  enforce the conservation of 
some  observable, quadratic in the velocity via, for instance, Gauss's 
principle for anholonomous constraints \cite{whittaker1988treatise,morriss2013statistical}.

The construction of the reversible models is not unique, relying on the 
choice of the observable to keep constant in the time-reversible 
dynamics. We found that the equivalence between the two statistical 
ensembles holds, as expected, in the quasiequilibrium regime, i.e., in 
the limit of very large Reynolds number when keeping constant the 
number of shells (i.e., the ultraviolet cutoff). Moreover,   
when the reversible model is constructed by imposing a constraint 
impacting preferentially the smallest and fastest scales of the system, 
e.g., when enforcing the conservation of enstrophy, equivalence 
is obtained  also in the energy-cascade regime, likely, owing to the 
robustness of the cascade mechanisms against the mechanism of energy 
dissipation.

The results in this study, together with similar findings for the 2D 
Navier-Stokes equations \cite{gallavotti2004lyapunov} and the Lorenz 
system \cite{gallavotti2014equivalence}, strengthen the case for 
the nonequilibrium statistical equivalence to hold also for other 
physically relevant nonequilibrium dynamical systems and in 
particular for the 3D Navier-Stokes equations, for which it was 
originally conjectured \cite{gallavotti1996equivalence}.

Besides the theoretical interest, the results here presented offer more 
freedom in modeling viscous forces in nonequilibrium systems, with 
particular reference to the ones of interest in fluid dynamics. 
Specifically, the ideas discussed in this paper could be relevant for 
small-scale  parametrization in atmosphere, ocean, and climate models 
\cite{Palmer2009,Franzke2015,Berner2017}, as well as LES models 
\cite{galperin1993large,sagaut2006large}, where eddy viscosity need to 
be carefully tailored in order to have results compatible with DNSs. 
Indeed, some form of reversible modeling of the small-scale dynamics is 
already used in LES  \cite{carati2001modelling,fang2012time}. 

\section*{Acknowledgments}
G.G. thanks L. Pizzochero for fruitful discussions and for suggesting
Ref.~\cite{CLT007}. L.B. and M.D.P. acknowledge funding from the European
Research Council under the European Union's Seventh Framework
Programme, ERC Grant Agreement No. 339032. L.B., M.C. and M.D.P. acknowledge
the European COST Action No. MP1305 “Flowing Matter.”  V.L. acknowledges the
support of DFG Sftb/Transregio Project No. TRR181.

\appendix
\section{Numerical integration scheme}
\label{sec:appendix_A}
Equations 
(\ref{eq:sabra_standard_equations})--(\ref{eq:sabra_reversible_generic_equations}), 
neglecting the forcing term, have the structure
\begin{equation}
\label{eq:shellmodel_equations_structure}
\frac{d}{dt} u_n(t) = g_n[\{u_n(t)\}] - \nu k_n^2 u_n(t) \, ,
\end{equation}
where $g_n[\{u_n(t)\}]$ stands for the nonlinear term at shell $n$, 
calculated on the velocity configuration $\{u_n(t)\}$ at time $t$.

When $\nu$ is constant in time, we adopted the following modified fourth 
order Runge-Kutta scheme, which exactly integrates the viscous 
contribution:
\begin{equation}
\label{eq:rk4exp_integration_scheme}
\begin{cases}
\vspace{0.2cm}
u_n(t + \delta t) = e_n \left\{ 
e_n \left[ 
u_n(t) + \frac{\delta t}{6} g_n[\{u_n(t)\}] \right] \right.\\ 
\vspace{0.2cm} \quad \left. + \frac{\delta t}{6} \left( g_n[\{u_n^{(1)}(t)\}] + g_n[\{u_n^{(2)}(t)\}] \right) + \frac{\delta t}{6} g_n[\{u_n^{(3)}(t)\}] \right\} \, , \\ 
\vspace{0.2cm}
u_n^{(1)}(t) = e_n \left[ u_n(t) + \frac{\delta t}{2} g_n[\{u_n(t)\}] \right] \, , \\  \vspace{0.2cm}
u_n^{(2)}(t) = e_n u_n(t) + \frac{\delta t}{2} g_n[\{u_n^{(1)}(t)\}] \, , \\  \vspace{0.2cm}
u_n^{(3)}(t) = e_n \left[ e^{\nu k_n^2 \delta t/2} u_n(t) + \frac{\delta t}{2} g_n[\{u_n^{(2)}(t)\}] \right] \, , \\  \vspace{0.2cm}
e_n = e^{\nu k_n^2 \delta t/2}  \, .
\end{cases}
\end{equation}
For the reversible models, where $\nu$ is not a constant, we introduced 
the following correction to the scheme:
\begin{equation}
\label{eq:rk4exp_integration_scheme_reversible}
\begin{cases}
\vspace{0.2cm}
u_n(t + \delta t) = e_n \left\{ e_n \left[ u_n(t) + \frac{\delta t}{6} \hat{g}_n[\{u_n(t)\}] \right] \right. \\ 
\vspace{0.2cm} \quad  \left. + \frac{\delta t}{6} \left( \hat{g}_n[\{u_n^{(1)}(t)\}] + \hat{g}_n[\{u_n^{(2)}(t)\}] \right) + \frac{\delta t}{6} \hat{g}_n[\{u_n^{(3)}(t)\}] \right\} \, , \\ \vspace{0.2cm}
u_n^{(1)}(t) = e_n \left[ u_n(t) + \frac{\delta t}{2} \hat{g}_n[\{u_n(t)\}] \right] \, , \\  \vspace{0.2cm}
u_n^{(2)}(t) = e_n u_n(t) + \frac{\delta t}{2} \hat{g}_n[\{u_n^{(1)}(t)\}] \, , \\  \vspace{0.2cm}
u_n^{(3)}(t) = e_n \left[ e^{\nu k_n^2 \delta t/2} u_n(t) + \frac{\delta t}{2} \hat{g}_n[\{u_n^{(2)}(t)\}] \right] \, , \\  \vspace{0.2cm}
e_n = e^{ \nu[\{u_n(t)\}] k_n^2 \delta t/2}  \, , \\  \vspace{0.2cm}
\hat{g}_n[\{u_n^{(i)}(t)\}] = g_n[\{u_n^{(i)}(t)\}] - (\nu[\{u_n^{(i)}(t)\}] \\ \quad - \nu[\{u_n(t)\}]) \, k_n^2 \, u_n(t)\, .
\end{cases}
\end{equation}


\end{document}